\documentclass[12pt]{article}

\usepackage{amssymb}

\begin{document}

\begin{titlepage}

\begin{flushright}
\end{flushright}
\vskip 2.5cm

\begin{center}
{\Large \bf No Contact Terms for the Magnetic Field in\\
Lorentz- and CPT-Violating Electrodynamics}
\end{center}

\vspace{1ex}

\begin{center}
{\large Karl Schober and Brett Altschul\footnote{{\tt baltschu@physics.sc.edu}}}

\vspace{5mm}
{\sl Department of Physics and Astronomy} \\
{\sl University of South Carolina} \\
{\sl Columbia, SC 29208} \\
\end{center}

\vspace{2.5ex}

\medskip

\centerline {\bf Abstract}

\bigskip

In a Lorentz- and CPT-violating modification of electrodynamics, the fields
of a moving charge are known to have
unusual singularities. This raises the question of whether the
singular behavior may include $\delta$-function contact terms, similar to
those that appear in the fields of idealized dipoles. However, by calculating the
magnetic field of an infinite straight wire in this theory,
we demonstrate that there are
no such contact terms in the magnetic field of a moving point charge.

\bigskip

\end{titlepage}

\newpage

\section{Introduction}

Much of the history of modern physics has involved symmetries that initially
appeared to be exact, yet which are actually violated in subtle fashions.
Physics beyond the standard model might involve new forms of symmetry breaking.
Among the most extreme symmetry violations that might occur in new physics
are the breaking of Lorentz and CPT symmetries. These symmetries are related
to isotropy, boost invariance, and hermiticity of the Hamiltonian. These
features underlie both the standard model and general relativity, but
quantum gravity theories could be different. In fact, many schematic theories
of quantum gravity appear to have regimes in which Lorentz and CPT symmetries
do not hold. Conversely, if evidence of these kinds of fundamental
symmetry violations were ever to be uncovered, that would provide
powerful evidence about the shape of new physics beyond what we currently
understand.

Exotic theories with unusual characteristics can also provide interesting
theoretical laboratories for understanding the general structure of quantum
field theories. Even if Lorentz and CPT symmetries are exact in nature,
such theories may provide fundamental insights about the kinds of behaviors that
are permitted in general field theories. The natural formalism for approaching
these kinds of problems is effective field theory. The effective field theory
that describes Lorentz and CPT violation is known as the standard model
extension (SME), and it has been the subject of extensive study.
The SME action is constructed from all operators that may be built up from
standard model fields~\cite{ref-kost1,ref-kost2}. Without the requirement
of Lorentz invariance, the number of possible operators is exceedingly large.
For practical calculations, a standard theory for discussing these broken
symmetries is known as the minimal SME; this is the subset of the SME that
contains only the finite number of operators that are local, are power counting
renormalizable, and respect the gauge symmetries of the standard model.
Most experimental bounds on Lorentz violations are formulated in terms of
constraints on minimal SME operators.

Some forms of Lorentz and CPT violation have more peculiar properties than others.
Possibly the most unusual terms in the minimal SME have what is known as
a Chern-Simons form.
The electromagnetic Cherm-Simons term affects the propagation of left- and
right-handed photons differently. At relatively short wavelengths, the
differences between the two modes' dispersion relations lead to a polarization
rotation during propagation. At sufficiently long wavelengths, the frequency
for one of the modes may become imaginary, signaling an instability. These and
other unconventional features make the Chern-Simons theory particularly
interesting as a tool for understanding how novel quantum field theories may
potentially behave.

Because the electromagnetic Chern-Simons term breaks parity and CPT symmetries, the left-right
asymmetry in wave propagation speeds would lead to photon birefringence.
The distinctive birefringence signature has been searched for and not found,
even for waves coming from sources at cosmological distances~\cite{ref-carroll1,ref-kost21,ref-mewes5}.
The lack of birefringence has been used to place exceedingly tight bounds on
the coefficient of the real-world Chern-Simons term. Nevertheless, the Chern-Simons
theory is still of theoretical interest, because the theory has some very unusual
features. For example, the Chern-Simons Lagrange density is not gauge invariant; it
changes, but only by a total derivative, under a gauge transformation. This fact makes the
determination of the radiative corrections to the Chern-Simons term a very subtle
problem, and the topic led to a significant amount of controversy~\cite{ref-coleman,
ref-jackiw1,ref-victoria1,ref-chung1,ref-andrianov,ref-altschul1}.

We have previously investigated another peculiar feature in this theory---the
possibility of vacuum Cerenkov radiation. Cerenkov processes are normally forbidden in
vacuum by Lorentz invariance, but with Lorentz violation, the phase speed of light need
not be uniformly 1 for all directions and frequencies.
Because the Chern-Simons term affects the dispersion relation for propagating waves,
radiation by charges in uniform motion (no matter how slow) becomes kinematically
allowed in the theory. However,
our investigations have showed that in the case of a timelike Chern-Simons parameter,
there is no net radiation loss from a moving charge~\cite{ref-schober}

In the course of our investigations, we developed an iterative algorithm for
determining the electric and magnetic fields of a moving point charge. The
geometry of the solution is incompatible with radiation emission, but the
structure of the fields is quite unusual, with singularities rather unlike
those seen in the conventional electrodynamics of point sources. Since in the
standard Maxwell theory, the fields of point dipoles include $\delta$-function
contact terms, it is natural to wonder whether there are analogous contact terms
in the Chern-Simons theory.

This paper continues the analysis of the Lorentz-violating Chern-Simons theory.
In section~\ref{sec-theory}, we describe the theory in detail and point out
some of the peculiar features of the solutions that we have previously uncovered.
In section~\ref{sec-wire}, we solve for the magnetostatic field of an infinite straight
wire in the modified theory. This solution is somewhat interesting on its own, and it
will also be possible to relate the new solution to the question of contact terms.
Our conclusions about the contact term question are presented in section~\ref{sec-concl}.

\section{Lorentz-Violating Chern-Simons Electrodynamics}
\label{sec-theory}

The Lagrange density for the photon sector of the minimal SME is
\begin{equation}
{\cal L}= -\frac{1}{4}F^{\mu\nu}F_{\mu\nu}
-\frac{1}{4}k_{F}^{\mu\nu\rho\sigma} F_{\mu\nu}F_{\rho\sigma}
+\frac{1}{2}k_{AF}^{\mu}\epsilon_{\mu\nu\rho\sigma}F^{\nu\rho}A^{\sigma}
-j^{\mu}A_{\mu}.
\end{equation}
In addition to the usual photon propagation term and the interaction with
charged matter, there are two operators with tensor-valued coefficients.
These are the sources of Lorentz violation. The Chern-Simons term is the $k_{AF}$ term,
and it is odd under CPT. The $k_{F}$ term is CPT even and has also been the
subject of extensive study. However, it is not our focus here, and we shall
neglect it. We are also interested only in the case of the timelike Chern-Simons
term, for which there is a frame in which $k_{AF}^{\mu}=(k,\vec{0}\,)$. There
are many important differences between the spacelike and timelike cases.

The dispersion relations for right- and left-circularly polarized waves are different
in the Chern-Simons theory. They take the forms $\omega_{\pm}^{2}=p(p\mp 2k)$. At very
long wavelengths, one of the frequencies becomes imaginary. This signals an
instability in the theory. The energy is not bounded below, and there are 
runaway solutions with exponentially growing field amplitudes. Moreover, it
is not possible to simply excise these modes from the theory; they play a key
role in the energy transport in the theory~\cite{ref-schober}. The runaway
solutions can be avoided through the use of an acausal Green's
function~\cite{ref-carroll1}, but this has the obviously problematic feature that
charges begin radiating before they start moving.

The interest in Cerenkov radiation in this theory arose from the fact that, even
apart from modes with $\omega^{2}<0$, there are propagating modes in the theory with
arbitrarily small phase speeds. Any charge $q$ moving with speed $v$ will manage to
outpace some of these modes. Radiation into these modes is kinematically allowed,
although it turns out that a charge in uniform motion does not actually radiate
energy.

In our previous work~\cite{ref-schober}, we avoided questions about causality and
temporal boundary conditions by using a steady-state source configuration, with a single
charge moving at a constant velocity. The steady-state condition ensured that the only
time dependence was that the fields were carried along at the same velocity $v$ as the
charge. In our calculations in section~\ref{sec-wire}, we will avoid the possible problems
with time dependence
by using an even simpler configuration of sources, with a steady current flowing in an
infinite wire.

Once the difficulties associated with any time dependences are dealt with,
we may solve the modified Maxwell's equations directly. The
equations are
\begin{eqnarray}
\vec{\nabla}\cdot\vec{E} & = & \rho \\
\vec{\nabla}\times\vec{E} & = & -\frac{\partial B}{\partial t}\\
\vec{\nabla}\cdot\vec{B} & = & 0 \\
\vec{\nabla}\times\vec{B} & = & \frac{\partial\vec{E}}{\partial t}+2k\vec{B}+\vec{J}.
\end{eqnarray}
Note that only the Ampere-Maxwell law is modified (when $k_{AF}$ is purely timelike),
and that in a steady-state system
with $\rho=0$, the electric field vanishes. In particular, in vacuum, we may take
the curl of $\vec{\nabla}\times\vec{B}=2k\vec{B}$, to get the Helmhotz equation
\begin{equation}
\label{eq-helmh}
\left[\vec{\nabla}^{2}+(2k)^{2}\right]\vec{B}=0.
\end{equation}

Using the modified Maxwell's equations in~\cite{ref-schober,ref-altschul36}, we found the
first few terms in the $\vec{E}$ and $\vec{B}$
of a moving charge, when the fields are expressed as a
power series in the Chern-Simons parameter $k$. The lowest-order $k$-dependent term (at
the instant when the charge $q$ moving with velocity $v\hat{z}$ is located at the origin) is
\begin{equation}
\label{eq-B11}
\vec{B}^{(1,1)}=\frac{kqv}{4\pi r}(2\cos\theta\hat{r}-\sin\theta\hat{\theta}).
\end{equation}
[The indices on $\vec{B}^{(1,1)}$ indicate that it is first order in $k$ and first order in $v$.]
This field has a singularity at $r=0$, as do the conventional fields, but the structure is
somewhat different.

The field terms involving higher powers of $k$ are progressively less
singular at the location of the charge. This is just an outgrowth of the fact that essentially
the only dimensionless parameter in the problem of a uniformly moving charge (with no recoil)
is $kr$. So increasing powers of $k$ must be accompanied by increasing powers of $r$. This makes
terms with higher powers of $k$ better behaved at $r=0$; terms of ${\cal O}(k^{3})$ and higher
are found to be regular at the origin. Conversely, the fields, taken term by term, grow
increasingly quickly at large $r$. The symmetry properties of the fields dictate that there
is no radiation emission at infinity, but it is unclear whether the separate fields really
grow large for $r\gg 1/k$, or whether they can be resummed into a bounded function---although
the latter possibility seems more likely.

The fields we will derive in section~\ref{sec-wire} differ from the fields of a single moving
point charge in a couple significant ways. Firstly, the simple power counting arguments
relating power of $k$ to powers of the radius will
break down. With an infinite wire source, there are fields that depend on $\ln(k\rho)$, and the
new singularity structure will complicate things. Secondly, it will be clear
[from (\ref{eq-helmh})] that when the source is
static, the terms we find by expanding in powers of $k$ must sum to a function that decays
as $r\rightarrow\infty$.

For comparison with $\vec{B}^{(1,1)}$, in conventional electrodynamics the magnetic field of a
pointlike dipole $\vec{m}=m\hat{z}$ is
\begin{equation}
\label{eq-Bdip}
\vec{B}_{{\rm dip}}=\frac{m}{4\pi r^{3}}(2\cos\theta\hat{r}+\sin\theta\hat{\theta})+
\frac{2}{3}m\hat{z}\delta^{3}(\vec{r}\,).
\end{equation}
The $\delta$-function is critically important in some applications.
For example, it is responsible for
the bulk of the hyperfine splitting in the $S$ states of atomic hydrogen. (Physically, the
hyperfine interaction is dominated by the time that the electron spends inside the nucleus.)
The angular structures of $\vec{B}_{{\rm dip}}$ and $\vec{B}^{(1,1)}$ are seemingly similar.
That raises the natural question of whether (\ref{eq-B11}) is really correct, or whether that
field may also include a contact term at $r=0$.

This is the central question that this paper
will address: whether there are any analogous contact terms in the magnetic field of a moving
charge in the Chern-Simons theory. Note that a $\delta$-function term is not ruled out by simple
dimensional analysis. The dimensions
of the two terms in (\ref{eq-Bdip}) match because $1/r^{3}$ and
$\delta^{3}(\vec{r}\,)$ have the same units. On the other hand (\ref{eq-B11}) is proportional to
$k/r$, which has the same units as $\delta^{3}(\vec{r}\,)/k$. So if there are three-dimensional
$\delta$-function contact terms in the field in the Chern-Simons theory, they may involve
negative powers of $k$; and although our calculation method will begin with an expansion in
powers of $k$, we shall see that the method is indeed capable of finding fields that are not
analytic functions of $k$.

A standard way of deriving the dipole term in conventional magnetostatics is to take the
average of $\vec{B}$ over a spherical region. The magnetic sources inside the sphere
make a contribution to the average that is proportional to their total dipole moment. However,
in the Chern-Simons theory, the sources of $\vec{B}$ include $2k\vec{B}$ itself, and the
standard technique is not so useful.

A different strategy is required.
In order to simplify our search for evidence of contact terms, we shall relate the field of
the moving charge to the field of another source configuration---an
infinite current-carrying filamentary wire. The field of such a wire
is a problem of interest on its own, since
this is one of the standard idealized configurations studied in magnetostatics.
In addition to providing a solution to the contact term question, the
calculation will also have some interesting features that help
provide a fuller understanding of how these Lorentz- and CPT-violating theories behave.

An ideal one-dimensional current-carrying wire is equivalent to two 
line charges $\lambda=\pm I/2v$ moving in opposite direction with velocities
$\pm v\hat{z}$. These lead to a net convective current $I\hat{z}$.
The virtue of this configuration is that the electric fields manifestly
cancel. Without the Chern-Simons term, the $\vec{E}$
fields generated by the opposite line charges
sum to zero. The only way $\vec{E}$ can be modified by the Chern-Simons term is
indirectly, through $\vec{B}$; however, $\vec{B}$ is time independent by design, so
there are no additional electric fields generated through Faraday's law. The fact that
$\vec{E}=0$ can also be demonstrated using the symmetry properties of the field that
were determined in~\cite{ref-schober}.

Of course, $\vec{B}$ is nonzero in this configuration. Obviously, there is an azimuthal
component $B_{\phi}$, and there may also be a nonzero $B_{z}$. However, a radial field
$B_{r}$ is incompatible with the symmetry of the theory under time reversal. Since $\hat{z}$,
the direction of the motion, is the only preferred direction in the problem of the moving
charge, a contact term must appear as a contribution $\tilde{B}_{z}=qA\delta^{3}(\vec{r}\,)$.
When integrated over the moving charges in the infinite wire, for which $dq=(I/v)dz$,
this will give a contribution to the field of the wire
\begin{equation}
\int_{-\infty}^{\infty}\left(\frac{I}{v}dz\right)A\delta^{3}(\vec{r}\,)=\frac{IA}{v}
\delta^{2}(\vec{\rho}\,),
\end{equation}
where $\vec{\rho}=\rho\hat{\rho}$ is the projection of the position vector into the
$xy$-place.
So a two-dimensional $\delta$-function would remain in the field of the wire. We shall now
determine the field of the wire and demonstrate that no such term appears.

\section{Magnetic Field of the Wire}
\label{sec-wire}

As we shall see, the behavior of $B_{z}$ for an infinitely thin wire is a bit complicated,
even without a $\delta$-function singularity. We shall therefore first consider
the problem of a wire with finite radius $a$. Ultimately, we shall take $a\rightarrow 0$;
this is analogous to letting the radius of a uniformly magnetized sphere go to zero to
derive (\ref{eq-Bdip}). The wire is
located along the $z$-axis, carrying a current $\vec{I} = I\hat{z}$, uniformly spread over
its interior. We now proceed to solve the modified steady-state magnetic equations
\begin{eqnarray}
\vec{\nabla}\cdot\vec{B} & = & 0 \\
\vec{\nabla}\times\vec{B} & = & 2k\vec{B}+\vec{J}
\end{eqnarray}
with this source.

We write the magnetic field as a series, with each term $\vec{B}^{(m)}$ in the sum proportional
to $k^{m}$, 
\begin{equation}
\vec{B} = \sum_{m=0}^{\infty}\vec{B}^{(m)}.
\end{equation}
When plugged into the field equations, this expansion gives
\begin{eqnarray}
\label{eq-curl-Bm}
\vec{\nabla}\times\vec{B}^{(m)} & = & \left\{
\begin{array}{ll}
\vec{J}, & m=0\\
2k\vec{B}^{(m-1)}, & m \geq 1
\end{array}\right. \\
\label{eq-div-Bm}
\vec{\nabla}\cdot\vec{B}^{(m) }& = & 0
\end{eqnarray}
From these, the
fields may be found iteratively, starting with the conventional field $\vec{B}^{(0)}$,
\begin{equation}
\label{eq-Bzero}
\vec{B}^{(0)} = \left\{\begin{array}{ll}
\frac{I\rho}{2\pi a^2} \hat{\phi}& \rho \leq a \\
\frac{I}{2\pi\rho} \hat{\phi} & \rho > a
\end{array}\right..
\end{equation}
[The analogue of the magnetostatic Biot-Savart law in this theory is an integral
equation for $\vec{B}$. Solving this equation iteratively for the $\vec{B}^{(m)}$ would
be equivalent to the pseudo-Amperean approach we shall take.]

The symmetries of the problem simplify the calculation of the other $\vec{B}^{(m)}$ terms
considerably. $\vec{B}^{(0)}$ is an azimuthal field; it points in the $\hat{\phi}$ direction
and its magnitude is independent of both $\phi$ and $z$. The curl of an azimuthal field is a
longitudinal field, which points in the $z$-direction and is also independent of $\phi$ and
$z$, and the curl of a longitudinal field is azimuthal. We have that $\vec{B}^{(0)}$ is an
azimuthal field, and the curl of $\vec{B}^{(1)}$ must then be azimuthal, so $\vec{B}^{(1)}$
is longitudinal. This argument can be continued by induction on $m$, and we obtain
\begin{equation}
\label{eq-split-fields}
\vec{B}^{(m)} = \left\{\begin{array}{ll}
B_{\phi}^{(m)}(\rho)\,\hat{\phi} & $m$\,{\rm even} \\
B_{z}^{(m)}(\rho)\,\hat{z} & $m$\,{\rm odd}.
\end{array}\right..
\end{equation}
Since there is now only a single term at each order, it will be possible to solve for
the $\vec{B}^{(m)}$ iteratively, using only pseudo-Amperean techniques.

However, it is actually possible to do much better, since we know that the field
$\vec{B}$ obeys (\ref{eq-helmh}) in vacuum (that is, for $\rho>a$).
In cylindrical coordinates, $\vec{B}$ outside of the wire should then be a linear
combination of Bessel functions. We write the general cylindrically symmetric solution
as
\begin{equation}
\label{eq-Bz}
B_{z}(\rho) = \alpha J_{0}(2k\rho)+\beta N_{0}(2k\rho),
\end{equation}
where $J_{0}$ and $N_{0}$ are Bessel functions of the first and second kind and.
$B_{\phi}(\rho)$ must then be 
\begin{equation}
\label{eq-Bphi}
B_{\phi}(\rho)=-\frac{1}{2k}\frac{dB_{z}}{d\rho}=\alpha J_{1}(2k\rho)+\beta N_{1}(2k\rho)
\end{equation}
Naturally, this is the general azimuthal solution of (\ref{eq-helmh}) that is symmetric
under rotations around $\hat{z}$ and translations along the $z$-axis.

To determine $\alpha$ and $\beta$, we calculate $\vec{B}^{(1)}$ and $\vec{B}^{(3)}$ directly and
compare the results to the exact solution (\ref{eq-Bz}). To find the fields iteratively, we use a
a pseudo-Amperean methodology. We consider a couterclockwise-oriented rectangular loop $R$ of length
$l$ in the $z$-direction and with sides parallel to $\hat{z}$ located at $\rho'=0$ and $\rho'=\rho$.
The modified Ampere-Maxwell law relates the integral of $\vec{B}^{(1)}$ around this loop to the
flux of $\vec{B}^{(0)}$ through it:
\begin{eqnarray}
\label{eq-ampere}
\int_{R}d\vec{l}\cdot\vec{B}^{(1)}(\rho') & = & 2k\int_{0}^{l}dz'\int_{0}^{\rho}d\rho'\,
\vec{B}^{(0)}(\rho')\cdot\hat{\phi} \\
l[B_{z}^{(1)}(0)- B_{z}^{(1)}(\rho)] & = & 2kl\left\{\begin{array}{ll}
\int_{0}^{\rho}d\rho'\frac{I\rho'}{2\pi a^{2}}, & \rho\leq a \\
\int_{0}^{a}d\rho'\frac{I\rho'}{2\pi a^2}+\int_{a}^{\rho}d\rho'\frac{I}{2\pi\rho'}
& \rho>a
\end{array}\right. \\
\label{eq-Bz1}
B_{z}^{(1)}(\rho) & = & \left\{\begin{array}{ll}
-\frac{kI\rho^{2}}{2\pi a^{2}}+B_{z}^{(1)}(0), & \rho \leq a \\
-\frac{kI}{\pi}\ln\frac{\rho}{a}-\frac{kI}{2\pi}+B_{z}^{(1)}(0), & \rho > a
\end{array}\right..
\end{eqnarray} 
Here, $B_{z}^{(1)}(0)$ is a constant, which must still be determined. The presence of this
on-axis field could easily be overlooked, but it will prove crucial in our calculations.
Obviously, this quantity is related to the singularity structure of the field on the axis,
which we are ultimately trying to determine. There is no constraint on the on-axis field
coming from any symmetry, nor can this field be determined using just pseudo-Amperean loop
techniques. The fields $B_{z}^{(m)}(0)$ for odd $m$ will all need to be determined by
some other method. On the other hand, there will be no analogous term to worry about in the
azimuthal field.

The $k$-dependent azimuthal field $\vec{B}^{(2)}$ may be found by a similar method, using a
pseudo-Amperean
loop $C$ with radius $\rho$, which lies parallel to the $xy$-plane and has its center at $(0,0,z)$,
\begin{equation}
\int_{C}d\vec{l}\cdot\vec{B}^{(2)}=2k\int_{0}^{\rho}d\rho'\,(2\pi\rho')B_{z}^{(1)}(\rho').
\end{equation}
Evaluating this gives
\begin{eqnarray}
B_{\phi}^{(2)}(\rho) & = & \left\{\begin{array}{ll}
\frac{2k}{\rho}\int_{0}^{\rho}d\rho'\,\rho'\left[-\frac{kI\rho'^{2}}{2\pi a^{2}}+B_{z}^{(1)}(0)\right],
& \rho\leq a \\
\frac{2k}{\rho}\int_{0}^{a}d\rho'\,\rho'\left[-\frac{kI\rho'^{2}}{2\pi a^{2}}+B_{z}^{(1)}(0)\right]
+\frac{2k}{\rho}\int_{a}^{\rho}d\rho'\,\rho'\left[-\frac{kI}{\pi}\ln\frac{\rho'}{a}-\frac{kI}{2\pi}+B_{z}^{(1)}(0)\right], & \rho>a
\end{array}\right. \\
\label{eq-Bphi2}
B_{\phi}^{(2)}(\rho) & = & \left\{\begin{array}{ll}
-\frac{k^{2}I\rho^{3}}{4\pi a^{2}}+(k\rho)B_{z}^{(1)}(0) & \rho\leq a \\
-\frac{k^{2}I\rho}{\pi}\ln\frac{\rho}{a}-\frac{k^{2}Ia^{2}}{4\pi\rho}+(k\rho)B_{z}^{(1)}(0) & \rho > a
\end{array}\right..
\end{eqnarray}
Finally, we may perform another integration similar to (\ref{eq-ampere}), with $\vec{B}^{(2)}$
replacing $\vec{B}^{(0)}$ as the source term, to obtain $\vec{B}^{(3)}$. The result is
\begin{equation}
\label{eq-Bz3}
B_{z}^{(3)}=\frac{5k^{3}Ia^{2}}{8\pi}-\frac{k^{3}I\rho^{2}}{2\pi}+
\frac{k^{3}I}{\pi}\left(\rho^{2}+\frac{1}{2}a^{2}\right)\ln\frac{\rho}{a}
-(k\rho)^{2}B_{z}^{(1)}(0) + B_{z}^{(3)}(0)
\end{equation}
for $\rho>a$. It is only necessary to calculate $B_{z}^{(3)}$ outside the wire, since that is where
(\ref{eq-Bz}) holds. The fields inside the wire are needed only to obtain higher order terms. Knowing
$B_{z}^{(1)}$ and $B_{z}^{(3)}$, it is now possible to determine the coefficients $\alpha$ and $\beta$
in (\ref{eq-Bz}).

The leading behaviors of the Bessel functions $J_{0}$ and $N_{0}$ are
\begin{eqnarray}
J_{0}(2k\rho) & = & 1-(k\rho)^{2}+{\cal O}(k^{4}) \\
\label{eq-neumann}
N_{0}(2k\rho) & = & \frac{2}{\pi}\left\{[\ln(k\rho)+\gamma]\left[1-(k\rho)^{2}\right]+(k\rho)^{2}\right\}
+{\cal O}(k^{4}),
\end{eqnarray}
where $\gamma$ in (\ref{eq-neumann}) is the Euler-Mascheroni constant.
These Bessel functions, $J_{0}(2k\rho)$ and $N_{0}(2k\rho)$, contain only terms with even powers of $k$
(as well as logarithms in the case of $N_{0}$). According to (\ref{eq-split-fields}), $B_{z}$ can
contain only terms with odd powers of $k$, so $\alpha$ and $\beta$ may only have terms with odd powers of
$k$. This means that only the terms up to order $k^{2}$ are needed from the Bessel functions in order
to determine $\alpha$ and $\beta$, since the highest order field that has been directly calculated is
$B_z^{(3)}$. The right-hand side of equation (\ref{eq-Bz}) may be expanded,
\begin{eqnarray}
\label{eq-Bz-complete}
\alpha J_{0}+\beta N_{0} & = & \alpha\left[1-(k\rho)^{2}\right]
+\beta\frac{2}{\pi}\left\{[\ln(k\rho)+\gamma]\left[1-(k\rho)^{2}\right]+(k\rho)^{2}\right\}+{\cal O}(k^{5}) \\
& = & \left(\alpha+\beta\frac{2\gamma}{\pi}\right)\left[1-(k\rho)^{2}\right]+\beta\frac{2}{\pi}(k\rho)^{2} \\
& & +\,\beta\frac{2}{\pi}\left[1-(k\rho)^{2}\right]\ln(k\rho)+{\cal O}(k^{5}). \nonumber
\end{eqnarray}

We need to compare this to the sum of the two lowest-order longitudinal field terms,
\begin{eqnarray}
\label{eq-Bz-partial}
B_{z}^{(1)}(\rho)+ B_{z}^{(3)}(\rho) & = & -\frac{kI}{\pi}\left[1-(k\rho)^{2}-\frac{(ka)^{2}}{2}\right]
\ln(k\rho)-\frac{kI}{2\pi}\left[1+(k\rho)^{2}-(ka)^{2}\right] \\
&  & +\left[1-(k\rho)^{2}\right]\left[B_{z}^{(1)}(0)+\frac{kI}{\pi}\ln(ka)\right]
+\left[B_{z}^{(3)}(0)+\frac{k^{3}Ia^{2}}{8\pi}-\frac{k^{3}Ia^{2}}{2\pi}\ln(ka)\right]. \nonumber 
\end{eqnarray}
The terms in (\ref{eq-Bz-partial}) have been grouped to show similarities with (\ref{eq-Bz-complete}).
In order to have equality between the two expressions, the terms in square brackets containing
$B_z^{(1)}(0)$ and $B_z^{(3)}(0)$ must vanish. This gives us a condition that determines the on-axis
fields $B_z^{(m)}(0)$. Once these terms are eliminated, the expression (\ref{eq-Bz-partial})
becomes
\begin{equation}
\label{eq-Bz-sum}
B_{z}(\rho)=-\frac{kI}{\pi}\left[1-\frac{(ka)^{2}}{2}\right]\left[1-(k\rho)^{2}\right]
\ln(k\rho)-\frac{kI}{2\pi}\left[1-(k\rho)^{2}\right]-\frac{kI}{\pi}\left[1-\frac{(ka)^{2}}{2}\right]
(k\rho)^{2}+{\cal O}(k^{5}).
\end{equation}
Equating (\ref{eq-Bz-complete}) and (\ref{eq-Bz-sum}) immediately yields equations for
$\alpha$ and $\beta$,
\begin{eqnarray}
\frac{2}{\pi}\beta & = & -\frac{kI}{\pi}\left[1-\frac{(ka)^{2}}{2}\right]+{\cal O}(k^{5}a^{4}) \\
\alpha+\frac{2\gamma}{\pi}\beta & = & -\frac{kI}{2\pi}+{\cal O}(k^{5}a^{4}),
\end{eqnarray}
with solutions
\begin{eqnarray}
\alpha & = & \frac{kI}{\pi}\left\{\gamma\left[1-\frac{(ka)^{2}}{2}\right]-\frac{1}{2}\right\}+
{\cal O}(k^{5}a^{4}) \\
\beta & = & -\frac{kI}{2}\left[1-\frac{(ka)^{2}}{2}\right]+{\cal O}(k^{5}a^{4}).
\end{eqnarray}
In addition to $\alpha$ and $\beta$, there is another quantity $B_{z}(0)$, which
characterizes the solution in the interior of the wire. We have also determined it up to
${\cal O}(k^{3})$, 
\begin{equation}
\label{eq-Bz0final}
B_{z}(0)=-\frac{kI}{\pi}\left\{\left[1-\frac{(ka)^{2}}{2}\right]\ln(ka)+\frac{(ka)^{2}}{8}\right\}
+{\cal O}(k^{5}a^{4}).
\end{equation}

It is important to note that the higher order terms in $\alpha$ and $\beta$ must contain only
powers of $ka$. This is true for dimensional reasons, since the only dependence on $\rho$ in
(\ref{eq-Bz}) must be inside the Bessel functions. Conversely, the Bessel functions
$J_0(2k\rho)$ and $N_0(2k\rho)$ do not have any dependence on $a$. Any dependence on $a$ must
be absorbed into $\alpha$, $\beta$, and $B_{z}(0)$.

Instead of using the longitudinal field $B_{z}$, we could have determined $\alpha$ and $\beta$
using (\ref{eq-Bphi}) and gotten the same results. To verify this, we note that the two Bessel
functions $J_{1}$ and $N_{1}$ may be expanded
\begin{eqnarray}
J_{1}(2k\rho) & = & k\rho+{\cal O}(k^{3}) \\
N_{1}(2k\rho) & = & -\frac{1}{\pi k\rho}+\frac{2}{\pi}(k\rho)\ln(k\rho)+\frac{k\rho}{\pi}
(2\gamma-1)+{\cal O}(k^{3}).
\end{eqnarray}
Using the now known values of $\alpha$ and $\beta$, (\ref{eq-Bphi}) becomes
\begin{equation}
\label{eq-J1N1}
\alpha J_{1}(2k\rho)+\beta N_{1}(2k\rho)=\frac{I}{2\pi\rho}-\frac{k^{2}I\rho}{\pi}\ln(k\rho)
-\frac{k^{2}Ia^{2}}{4\pi\rho}+{\cal O}(k^{4});
\end{equation}
the first term on the right-hand side of (\ref{eq-J1N1}) is just the usual $\vec{B}$ of
an infinite wire.
Alternatively, since $B_z^{(1)}(0)$ has also been determined, (\ref{eq-Bphi2}) and
(\ref{eq-Bz0final}) give
\begin{equation}
B_{\phi}^{(2)}=-\frac{k^{2}I\rho}{\pi}\ln(k\rho)-\frac{k^{2}Ia^{2}}{4\pi\rho}, 
\end{equation}
which clearly agrees with (\ref{eq-J1N1}).
Notice how the $\ln(\rho/a)$ term in (\ref{eq-Bphi2}) and $\ln(ka)$ from
(\ref{eq-Bz0final}) combine to produce a final $B_{\phi}^{(2)}$ that does not
depend on a logarithm of the wire radius $a$. In fact, the $\vec{B}_{z}(0)$
terms supply the logarithms of $k$ that are needed to match the behavior of the
$N_{j}(2k\rho)$. The power series expansion in $k$ was not capable of generating
$\ln(k\rho)$ terms directly; notice that this logarithm does not appear in
(\ref{eq-Bz1}), (\ref{eq-Bphi2}), or (\ref{eq-Bz3}).

We also note that we could have determined the leading order behavior of
$\alpha$ and $\beta$ without actually calculating $B_{z}^{(3)}$. This could have
been accomplished by using comparisons with both $B_{z}^{(1)}$ and $B_{\phi}^{(2)}$.
However, by calculating $B_{z}^{(3)}$ explicitly, we were able to show something
interesting---that
there actually are correction terms in $\alpha$ and $\beta$ that are suppressed
by higher powers of $ka$.

Finally, we may take the limit as $a\rightarrow 0$, with the wire becoming an
infinitely narrow filament.
Since any higher order terms in $\alpha$ and $\beta$ are proportional to $a$, these
terms disappear as $a\rightarrow 0$. What is left is
\begin{equation}
\label{eq-Bfinal}
\vec{B}=\left[\frac{kI}{\pi}\left(\gamma-\frac{1}{2}\right)J_{1}(2k\rho)
-\frac{kI}{2}N_{1}(2k\rho)\right]\hat{\phi}+\left[\frac{kI}{\pi}\left(\gamma-\frac{1}{2}\right)
J_{0}(2k\rho)-\frac{kI}{2}N_{0}(2k\rho)\right]\hat{z}.
\end{equation}
The Euler-Mascheroni constant $\gamma$ is present because of the way the $N_{j}$ are
defined; it does not actually enter the expressions for the field at any order in $k$.

\section{Conclusion}
\label{sec-concl}

At this point, it may still not be entirely obvious whether we have found a solution
with a contact term or not. The solution (\ref{eq-Bfinal}) for $\vec{B}$ arose as the
limit of expressions that had different behaviors inside and outside the wire. This
is qualitatively similar to what one obtains by considering a uniformly magnetized
sphere of radius $R$, with fixed $\vec{m}$ as $R\rightarrow 0$. This is another
way to derive the contact term for a pointlike dipole. The $\delta$-function
in (\ref{eq-Bdip}) arises from
the behavior of $\vec{B}$ in the $r<R$ region, which is radically different from the
exterior field.

However, there is no $\delta$-function term in this case.
To understand why this is so, it is worthwhile to remember why the contact term is
necessary in the standard magnetic field of a point dipole. The $\delta$-function in
(\ref{eq-Bdip})
ensures that the integral $\oint d\vec{S}\cdot\vec{B}=0$ over any closed surface---that
$\vec{B}$ is divergenceless, even where it is singular.
(In contrast, the different $\delta$-function in the electric field of a pointlike
electric dipole ensures that $\vec{E}$ is curl-free even at the source.)

The magnetic field in the vicinity of the moving charge in the Chern-Simons theory is
rather complicated, and it may not be obvious from inspection whether it is
divergence-free. On the other hand, the field of the wire---as the sum of a purely
longitudinal field at odd orders in $k$ and a purely azimuthal field at even
orders---unquestionably satisfies $\vec{\nabla}\cdot\vec{B}=0$. Considered
separately, the lines of the longitudinal field component all run parallel to the
$z$-axis; the lines of the azimuthal field are all circles. In either case, the
field lines do not terminate. So no correction term is needed to make the field
divergenceless even on the axis.

Of course, this only shows that a $\delta$-function is not needed to make the field
(\ref{eq-Bfinal}) divergenceless. It takes a bit more to demonstrate that
the $\delta$-function indeed does not exist.
For comparison, we first consider the field of a line of dipoles. With a dipole
moment $\tilde{m}$ per unit length, spread over a cylinder of radius $a$, the
magnetization is $\vec{M}=(\tilde{m}/\pi a^{2})\hat{z}$. The field is well known:
$\vec{B}=\vec{M}$ inside the cylinder, and $\vec{B}=0$ outside. The
$\delta$-function in the dipole field can be identified by looking at the
difference between the flux of $\vec{B}$ through a circle of radius $b>a$
and the flux of the external field extrapolated all the way down to $\rho=0$.
Since the external field vanishes, this flux difference is simply
$\Delta\Phi=\tilde{m}$, independent of $a$. This is precisely the hallmark of
two-dimensional $\delta$-function behavior.

For the field of the wire in the Chern-Simons theory, we may apply a similar
procedure. First, we calculate the flux of the field $\vec{B}^{(1)}$ at finite $a$
through the same circular surface of radius $b$. Only the lowest order term
$\vec{B}^{(1)}$ from (\ref{eq-Bz1}) is necessary for this calculation; terms with
higher powers of $k$ will also involve higher powers of $a$, and so will vanish
more rapidly as $a\rightarrow 0$. The integral using the field (\ref{eq-Bz1}) is
\begin{equation}
\int_{0}^{b}2\pi b\,db\,B_{z}^{(1)}(\rho)=-kIb^{2}\ln(kb)-\frac{kIa^{2}}{4}.
\end{equation}
If we take instead the exterior field (\ref{eq-Bfinal}),
\begin{eqnarray}
\int_{0}^{b}2\pi b\,db\,B_{z}(\rho) & = & bI\left(\gamma-\frac{1}{2}\right)J_{1}(2kb)
-\frac{\pi bI}{2}N_{1}(2kb)-\frac{I}{2k} \\
& = & -kIb^{2}\ln(kb)+{\cal O}(k^{3}).
\end{eqnarray}

The two fluxes differ only by terms that vanish as $a\rightarrow 0$.
If there were a $\delta$-function term in the field, it would yield a
$\Delta\Phi$ that would persist for any finite value of $a$. However,
we now know that no such term exists, and (\ref{eq-Bfinal}) is the full expression
for the field of an infinite filamentary wire. Returning then to our original
argument, the absence of a $\delta^{2}(\vec{\rho}\,)$ in (\ref{eq-Bfinal})
implies that there cannot be a $\delta^{3}(\vec{r}\,)$ term in the field of a
moving point charge.

This might not seem like a surprising result. In fact, it would be easy to
overlook the fact that the $\delta$-function might even be a
possibility. However, this result does reveal something new about the
structure of the Chern-Simons theory.

Moreover, we have, as part of our calculation, produced a field solution for
another source configuration. The calculation of $\vec{B}$ has a number of
interesting features---for example, the matching of the low-order terms,
derived using pseudo-Amerperan loop methods, with the exact static vacuum
solution in terms of Bessel functions. The Bessel function solutions
also show explicitly how the power series in $k\rho$ converges to a function
that decays at large $\rho$. It is still unclear whether the full fields
from~\cite{ref-schober} show similar decays at large $r$; that situation is more
complicated, with both $\vec{E}$ and $\vec{B}$ nonzero and each field having
$z$, $\rho$, and $\phi$ components. However, the result for the infinite wire
certainly suggests that something similar might happen with the field of a single
moving charge.

Of course, the infinite wire obviously represents an
unphysical situation, but under appropriate circumstances, it could be a good
approximation. As in conventional electrodynamics, the infinite wire
approximation would be useful in the Chern-Simons theory if the distance from
a wire were small compared to the wire's radius of curvature. For the
$k$-dependent corrections to be of meaningful size, of course, $1/k$ must be
small compared with the radius of curvature as well. Given the experimental
bounds on $k$, this is never going to be a realizable configuration, but the
Chern-Simons theory does teach us interesting things about the nature of
Lorentz-violating field theories.


\begin{thebibliography}{99}

\bibitem{ref-kost1}D. Colladay, V. A. Kosteleck\'{y}, Phys. Rev. D {\bf 55},
6760 (1997).
\bibitem{ref-kost2}D. Colladay, V. A. Kosteleck\'{y}, Phys. Rev. D {\bf 58},
116002 (1998).
\bibitem{ref-carroll1}S. M. Carroll, G. B. Field, R. Jackiw, Phys. Rev. D
{\bf 41}, 1231 (1990).
\bibitem{ref-kost21}V. A. Kosteleck\'{y}, M. Mewes, Phys. Rev. Lett. {\bf 97}, 140401
(2006).
\bibitem{ref-mewes5}V. A. Kosteleck\'{y}, M. Mewes, Phys. Rev. Lett.
{\bf 99}, 011601 (2007).
\bibitem{ref-coleman}S. Coleman, S. L. Glashow, Phys. Rev. D {\bf 59}, 116008
(1999).
\bibitem{ref-jackiw1}R. Jackiw, V. A. Kosteleck\'{y}, Phys. Rev. Lett.
{\bf 82}, 3572 (1999).
\bibitem{ref-victoria1}M. P\'{e}rez-Victoria, Phys. Rev. Lett. {\bf 83}, 2518
(1999).
\bibitem{ref-chung1}J. M. Chung, Phys. Lett. B {\bf 461}, 138 (1999).
\bibitem{ref-andrianov}A. A. Andrianov, P. Giacconi, R. Soldati, JHEP
{\bf 02}, 030 (2002).
\bibitem{ref-altschul1}B. Altschul, Phys. Rev. D {\bf 69}, 125009 (2004).
\bibitem{ref-schober}K. Schober, B. Altschul, Phys. Rev.D {\bf 92}, 125016 (2015).
\bibitem{ref-altschul36}B. Altschul, Phys. Rev. D {\bf 90}, 021701 (R) (2014).

\end{thebibliography}
\end{document}